# The First Light Curve Solutions and Period Study of BQ Ari

A. Poro[1,2], F. Davoudi[1], F. Alicavus[3,4], S. Khakpash[5], O. Basturk[6], E. M. Esmer[6], E. Lashgari[1], J. Rahimi[1], Y. Aladağ[8], N. Aksaker[7,8], A. Boudesh[1], M. Ghanbarzadehchaleshtori[1], A. Akyüz[8,9], S. Modarres[1], A. Sojoudizadeh[1], M. Tekeş[8], A. Solmaz[8]

[1]The International Occultation Timing Association Middle East section, Iran, info@iota-me.com
[2]Astronomy Department of the Raderon Lab., Burnaby, BC, Canada
[3]Çanakkale Onsekiz Mart University, Faculty of Arts and Sciences, Department of Physics, 17020, Çanakkale, Turkey
[4]Çanakkale Onsekiz Mart University, Astrophysics Research Center and Ulupınar Observatory, 17020, Çanakkale, Turkey
[5]Department of Physics and Astronomy, University of Delaware, Newark, DE 19716, USA
[6]Ankara University, Faculty of Science, Astronomy and Space Sciences Department, TR-06100, Tandogan, Ankara, Turkey
[7]Adana Organised Industrial Zones Vocational School of Technical Science, University of Çukurova, 01410, Adana, Turkey
[8]Space Science and Solar Energy Research and Application Center (UZAYMER), University of Çukurova, 01330, Adana, Turkey
[9]Department of Physics, University of Çukurova, 01330, Adana, Turkey

**Abstract**

The first analysis of the photometric observation in BVR filters of a W UMa type binary system BQ Ari was performed. Light curve analysis was performed using Wilson-Devinney (W-D) code combined with a Monte Carlo (MC) simulation to determine its photometric and geometric elements and their uncertainties. These results show that BQ Ari is a contact binary system with a photometric mass ratio $q = 0.548 \pm 0.019$, a fillout factor $f = 24 \pm 0.8\%$, and an orbital inclination of $i = 85.09 \pm 0.45$. We used the parallax from Gaia EDR3 for calculating the absolute parameters of the binary system. This study suggested a new linear ephemeris for BQ Ari, combining our new mid-eclipse times and the previous observations, which we analyzed using the Monte Carlo Markov Chain (MCMC) method. We present the first analysis of the system's orbital period behavior by analyzing the O-C diagram using the Genetic Algorithm (GA) and the MCMC approaches in OCFit code. We attempted to explain the analysis of the residuals of linear fit in the O-C diagram with two approaches; "LiTE + Quadratic" and "Magnetic activity + Quadratic". Although we consider the magnetic activity to be probable, the system should be studied further in order to reveal the nature of orbital period variations.

## 1. INTRODUCTION

W UMa eclipsing variables may go through various physical phenomena throughout their lifetimes. Such phenomena are observed and studied to clarify why and how the components come in contact with each other. Therefore, analyzing this system's light curves while obtaining physical and geometrical parameters can explain their evolutionary process. Additionally, studies of orbital variations observed in contact binaries provide us with invaluable details about the mechanisms of mass transfer or mass loss of these systems.

We selected BQ Ari as an example of a W UMa binary for extensive analysis since it has a short period, and no photometric solution of the system has been published yet. BQ Ari is a W UMa type eclipsing binary star in the constellation of Aries. The variability of BQ Ari (GSC 00646-00946) was first discovered in the All-Sky Automated Survey (ASAS) project and then classified as a contact binary system. BQ Ari has an orbital period of 0.282333 days (Paschke 2011). Several ground observations of this system since 2011 have been made solely based on measurements of light curve minima. This study purseud two main goals. First of all, we present the first light curve analysis of BQ Ari intending to determine its geometrical and physical characteristics. Secondly, we examine changes in the system's orbital period over the years of observation using an O-C diagram to detect Eclipse Timing Variation (ETV), and we investigate different models that could explain these variations.

## 2. OBSERVATION AND DATA REDUCTION

The photometric observations of BQ Ari were carried out with a 50 cm Ritchey Chretien telescope and Apogee Aspen CG type CCD during four nights of observation at the UZAYMER Observatory Çukurova University, Adana, Turkey on the 7th and 15th of December 2019 and the 13th and 15th of January 2020. The CCD attached to the telescope has a 1024×1024 pixel array with a pixel length of 24μ. In these observations we used $BVR$ standard Johnson filters. Each of the frames was 1×1 binned with a 40s exposure time with an $R$ filter, 60s for a $V$ filter, and 85s for a $B$ filter; the average temperature of the CCD was -45°C during four nights of observation. A total of 1281 images were acquired with $BVR$ filters. Observations through $V$ and $R$ filters were performed for each filter in one single night and with the $B$ filter during two other nights.

BQ Ari belongs to the triple system[1]. The southern and brighter component is 14-arcsecond far from another component. These components are separate in our CCD images. This study observed a brighter component that is an eclipsing binary system, and this star's primary information is listed in Table 1. The fainter component is a northern star (02 48 41.03, +13 45 1.4, Gaia DR2 J2000)[2], and this star's temperature, according to Gaia DR2, is $T = 4892$ K.

GSC 646-868 and TYC 646-333-1 were chosen as comparison stars, and GSC 646-726 was selected as the reference star. All of these stars are close to BQ Ari, and the magnitude of the reference star is appropriate for precise differential photometry. The characteristics of the comparisons and reference stars are shown in Table 1.

Table 1. Characteristics of the variable, comparison and reference stars.

| Type | Star | Magnitude ($V$) | RA. (J2000) | Dec. (J2000) |
|---|---|---|---|---|
| Variable | BQ Ari | 10.76 | 02 48 40.73 | +13 44 48.02 |
| Comparison$_1$ | GSC 646-868 | 12.77 | 02 48 36.48 | +13 43 28.69 |
| Comparison$_2$ | TYC 646-333-1 | 10.93 | 02 48 27.00 | +13 44 58.73 |
| Reference | GSC 646-726 | 12.56 | 02 48 34.14 | +13 43 53.17 |

We reduced the raw images and corrected them. According to the standard method, the basic data reduction was performed for bias, dark and flat field of each CCD image. We aligned, reduced, and plotted raw images with the AstroImageJ (AIJ) software (Collins et al. 2017).

## 3. LIGHT CURVE ANALYSIS

We analyzed our light curves using the Wilson & Devinney (1971) code (W-D) combined with the MC simulation to determine the uncertainties of the adjustable parameters (Zola et al. 2004, 2010).

We used Gaia DR2 to find the primary component's temperature and fixed it in the light curve solutions. Based on our data and after the required calibrations (Høg et al. 2000), we calculated $(B-V)_{BQ\ Ari} = 0^m.65 \pm 0.07$. Thus, the effective temperature of the primary component, $T_1$ was assumed as $5559 \pm 208$ K (Eker et al. 2020). This temperature value is a good approximation because we can compare it with that from the Gaia DR2 catalogue which is $5498^{+25}_{-166}$ K; Based on the Gaia color BP-RP, the difference in temperature is consistent. This indicates that the accuracy of the observations is reliable.

We fixed some parameters and assumed gravity-darkening coefficients of $g_1 = g_2 = 0.32$ (Lucy 1967) and bolometric albedos of $A_1 = A_2 = 0.5$ (Rucinski 1969) for stars with convective envelopes; linear limb darkening coefficients were taken from the tables published by Van Hamme (1993). The parameters and their errors obtained from the light curve analysis are presented in Table 2. The synthetic light curves based on these parameters are given in Figure 1 from the models.

---

[1]http://simbad.u-strasbg.fr/simbad/
[2]https://www.cosmos.esa.int/web/gaia/dr2

Table 2. Photometric solutions of BQ Ari.

| Parameter | Results | Parameter | Results | Parameter | Results |
|---|---|---|---|---|---|
| $T_1$ (K) | 5498(200) | $f$ (%) | 24(8) | $r_1$(mean) | 0.449(5) |
| $T_2$ (K) | 5497(209) | $r_1$(back) | 0.4821(50) | $r_2$(mean) | 0.345(4) |
| $\Omega_1 = \Omega_2$ | 2.8877(299) | $r_1$(side) | 0.4479(46) | Colatitude$_{spot}$ (deg) | 20.5(1.3) |
| $i$ (deg) | 85.09(45) | $r_1$(pole) | 0.4197(43) | Longitude$_{spot}$ (deg) | 59.8(2) |
| $q$ | 0.548(19) | $r_2$(back) | 0.3787(39) | Radius$_{spot}$ (deg) | 48.8(1.7) |
| $l_1/l_{tot}(BVR)$ | 0.622(6) | $r_2$(side) | 0.3369(35) | $T_{spot}/T_{star}$ | 0.83(2) |
| $l_2/l_{tot}(BVR)$ | 0.378(4) | $r_2$(pole) | 0.3205(33) | Phase Shift | 0.0208(2) |

Notes: Parameters of a star spot on the primary component.

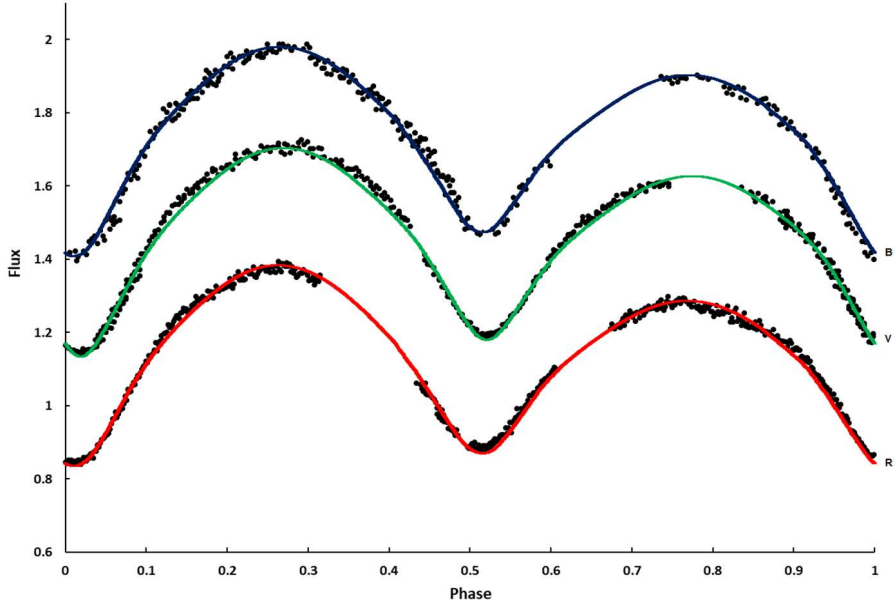

**Fig 1.** The observed light curves of BQ Ari (black dots), and synthetic light curves obtained from light curve solutions in the *B*, *V*, and *R* filters (top to bottom respectively); with respect to orbital phase, shifted arbitrarily in the relative flux.

To gain more reliable values for the absolute parameters of the binary system, we used the parallax from Gaia EDR3, the Early release of Gaia Data Release 3 (3 December 2020), and we calculated $d(pc) = 130.9408 \pm 2.3638$. Using our estimated value of apparent magnitude $V_{system} = 10^m.876 \pm 0.004$, and $A_{dV} = 0.177 \pm 0.021$ (Schlafly and Finkbeiner 2011), the value of absolute magnitude, $M_v$, was estimated. As our light curve solutions conclude that $l_1/l_2 = 0.6077$, we calculated $M_{v1}$ and $M_{v2}$. Applying $(BC) = -0.0824$ (Eker et. al 2020), we computed $M_{1bol}$ and $M_{2bo}$. Then, we calculated $L_1, L_2, R_1,$ and $R_2$ for the components of this binary system respectively.

The well-known relation $R = a \times r$ gives us $a_1$ and $a_2$, from $R_1$ and $R_2$, and also, we obtained the average of $a_1$ and $a_2$. The mean fractional radii of the components are 0.449(5) and 0.345(4) for the primary and secondary components, respectively; these values were calculated from the formula, $r_{mean} = (r_{pole} \times r_{side} \times r_{back})^{\frac{1}{3}}$. The binary system is in a marginal contact state (Kopal 1959) since the sum of the mean fractional radii of the components is $r_{mean} = r_{1me} + r_{2mean} = (0.79) > 0.75$. Therefore, we calculated the masses of the components, $M_1$ and $M_2$, by Kepler's third law. Finally, we can calculate $log(g)$ according to the relation between $M$ and $R$ values for each component. The parameters obtained during this procedure are given in Table 3.

Table 3. Estimated absolute elements of BQ Ari.

| Parameter | Primary | Secondary |
|---|---|---|
| $Mass$ ($M_\odot$) | 0.642(38) | 0.352(25) |
| $Radius$ ($R_\odot$) | 0.805(15) | 0.627(14) |
| $Luminosity$ ($L_\odot$) | 0.533(13) | 0.324(8) |
| $M_{bol}$ (mag.) | 5.42(29) | 5.96(36) |
| $M_v$ (mag.) | 5.54(15) | 6.08(17) |
| $log\ g$ (cgs) | 4.434(15) | 4.390(18) |
| $a$ ($R_\odot$) | 1.8054(17) | |

One of W UMa binaries' interesting characteristics is the well-known O'Connell effect (O'Connell 1951) that is described by the asymmetry in the brightness of maxima in the light curve of eclipsing binary star systems. The most appropriate suggestion for this effect is the presence of star spot(s) induced by the components' magnetic activities (Sriram et al. 2017).

The light curves of BQ Ari in the $B$, $V$, and $R$ bands (Figure 1) indicate the presence of the O'Connell effect in this system, whereby $Max\ I$ is brighter than $Max\ II$ ($Max\ I > Max\ II$) and asymmetry in maxima or unequal minima is clearly visible. Due to the presence of this asymmetry we used a stellar spot model during the light curve solution for all three $BVR$ filters. We found that assuming a cool spot model on the massive primary component results in an acceptable solution for all $BVR$ light curves (Table 2).

Table 4 represents BQ Ari light curves' characteristic parameters where the difference in maxima in each filter is shown in the first row. Accordingly, the largest difference between the maximum light levels and the depths of the minima were observed in the $B$ filter, whereas they were the smallest in the $R$ filter. This is expected from the magnetic activity-induced variations.

Table 4. Characteristic parameters of the light curves in $BVR$ filters.

| Light curve | $\Delta B$ | $\Delta V$ | $\Delta R$ |
|---|---|---|---|
| $MaxI - MaxII$ | -0.097 | -0.088 | -0.082 |
| $MaxI - MinII$ | -0.794 | -0.801 | -0.552 |
| $MaxI - MinI$ | -0.956 | -0.909 | -0.600 |
| $MinI - MinII$ | 0.162 | 0.108 | 0.048 |

The components' positions are shown in Figure 2 for four different phases during an orbital period. The cold spot is assumed to be on the primary star.

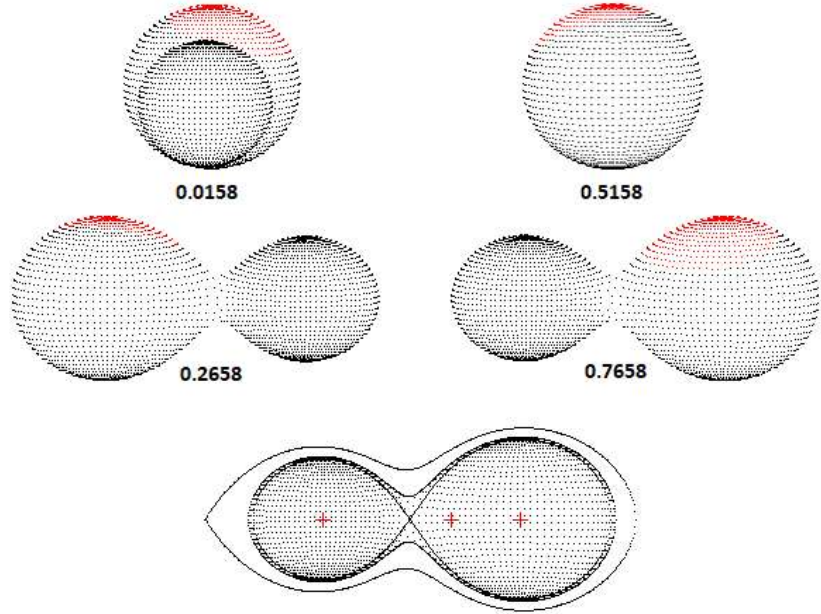

**Fig 2.** The positions of the components of BQ Ari.

**4. DETERMINATION OF MID-ECLIPSE TIMES BY MCMC**

Before analyzing period changes it is necessary to find the times of minima in the light curves. This is commonly achieved using the Kwee and van Woerden (1956) method (KW method) by computing the mid-eclipse times from the binary system's light curves. Despite it being a well-established method to derive minima times, the KW method underestimates the uncertainties, and the estimated errors are usually unrealistically small (Li et al. 2018; Pribulla et al. 2012). Furthermore, it may not be adequate in the case of asymmetric data or an incomplete light curve due to its nature (Mikulášek et al. 2014). To find the times of minima we fitted models based on Gaussian and Cauchy distributions to selected portions of the light curves that include the minima. We used the MCMC sampling methods to estimate the uncertainty of the values. The code for this part of the analysis is implemented in Python using the PyMC3 package (Salvatier et al. 2016).

**5. ORBITAL PERIOD VARIATIONS**

Formally, the O-C variations are shown by $\delta T$ and defined as

$$\delta T = (\Delta T_0 + \Delta P \times E) + Q \times E^2 + \delta T_i \qquad (1)$$

where the part in parentheses generates a linear trend in (O-C)s and it is caused by the uncertainties of light elements of the linear ephemeris. The quadratic term describes changes due to mass transfer in the binary system. The term $\delta T_i$ implies more complex periodic variations in (O-C)s induced by quasi-cyclic and cyclic physical behaviors (Gajdoš and Parimucha 2019).

We collected 17 mid-eclipse times from the literature and obtained 8 individual mid-eclipse times from our observations shown in Table 5. Mid-eclipse times identified as Min. (BJD$_{TDB}$), and these are in column 1; their uncertainties appear in column 2 (some observers did not provide their uncertainties, and we assumed the mean error value for these data points). Minima types (I: primary and II: secondary) are in column 3; epochs of these minima times are in column 4; O-C values are in column 5; and the references of mid-eclipse times are shown in the last column. The epochs and the O-C values were calculated by the linear ephemeris mentioned in Školník (2017).

Table 5. Times of minima of BQ Ari obtained by CCD.

| Min. (BJD$_{TDB}$) | Error | Epoch | O-C | Reference | Min. (BJD$_{TDB}$) | Error | Epoch | O-C | Reference |
|---|---|---|---|---|---|---|---|---|---|
| 2455593.3617 | 0.0020 | -8678 | -0.0010 | Paschke 2011 | 2458043.4658 | 0.0006 | 0.0 | 0.0000 | Školník 2017 |
| 2455866.3787 | 0.0041 | -7711 | -0.0019 | Hübscher et al. 2012 | 2458043.6081 | 0.0009 | 0.5 | 0.0011 | Školník 2017 |
| 2456563.4603 | 0.0009 | -5242 | -0.0054 | Hübscher 2014 | 2458438.3107 | 0.0041 | 1398.5 | -0.0006 | Pagel 2019 |
| 2456563.6023 | 0.0010 | -5241.5 | -0.0046 | Hübscher 2014 | 2458438.4518 | 0.0026 | 1399 | -0.0007 | Pagel 2019 |
| 2456573.4835 | 0.0004 | -5206.5 | -0.0051 | Zasche et al. 2014 | 2458825.2492 | 0.0002 | 2769 | -0.0022 | This study |
| 2456573.6258 | 0.0002 | -5206 | -0.0040 | Zasche et al. 2014 | 2458825.3909 | 0.0002 | 2769.5 | -0.0017 | This study |
| 2457320.2625 | 0.0041 | -2561.5 | -0.0022 | Nagai 2016 | 2458833.1547 | 0.0002 | 2797 | -0.0021 | This study |
| 2457331.4158 | 0.0030 | -2522 | -0.0011 | Paschke 2017 | 2458833.2958 | 0.0003 | 2797.5 | -0.0022 | This study |
| 2457331.5568 | 0.0040 | -2521.5 | -0.0013 | Paschke 2017 | 2458833.4372 | 0.0002 | 2798 | -0.0019 | This study |
| 2457385.3411 | 0.0017 | -2331 | -0.0018 | Hübscher 2017 | 2458862.2353 | 0.0003 | 2900 | -0.0020 | This study |
| 2457385.4831 | 0.0024 | -2330.5 | -0.0010 | Hübscher 2017 | 2458864.2119 | 0.0003 | 2907 | -0.0018 | This study |
| 2457657.5134 | 0.0001 | -1367 | -0.0005 | Pagel 2018 | 2458864.3531 | 0.0004 | 2907.5 | -0.0017 | This study |
| 2457728.2391 | 0.0002 | -1116.5 | 0.0003 | Ozavci et al. 2019 | | | | | |

We plotted the O-C diagram and fitted all mid-eclipse times with a line using a code that we wrote based on Python's emcee package. We used a least-squares approach to find initial values of $\Delta P$ and $\Delta T_0$ for the MCMC sampling, and determined a new linear ephemeris for the primary minimum. We used 200 random walkers in 10000 steps while the first 100 of each walker were ignored (burn-in = 100) to fix the chains at a certain value for slope and y-intercept of the line. In our modeling, the slope is the change in the period and the y-intercept is the change in the reference mid-eclipse time. Using this approach, the posterior probabilities for these parameters were obtained, and the resulting corner plot is shown in Figure 3. Consequently, we refined the ephemeris based on the combination of our mid-eclipse times and those from previous observations as

$$Min. I\ (BJD_{TDB}) = (2458043.4637775^{+0.0003128}_{-0.0003148}) + (0.28233519^{+0.00000008}_{-0.00000007}) \times E\ \text{days} \qquad (2)$$

where $E$ is the integer number of orbital cycles after the reference epoch.

After subtracting the linear trend from the original O-C, the residue (O-C) diagram (the residue between original values of this O-C and O-C's calculated from the fit) is plotted in Figure 4 concerning new linear ephemeris (Equation 2).

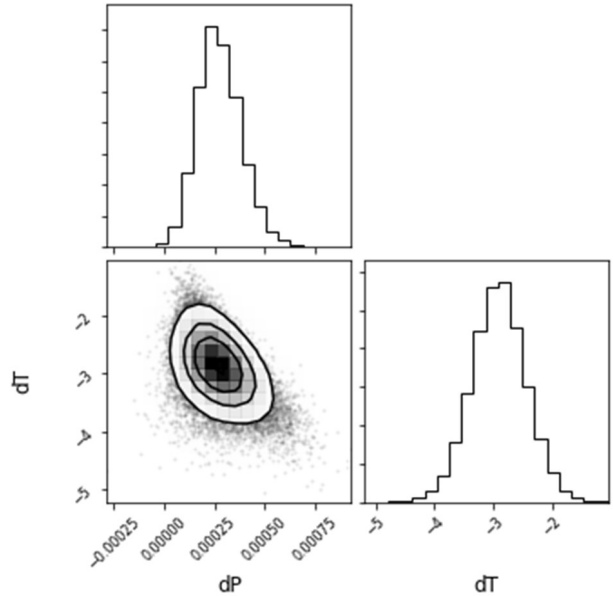

**Fig 3.** Corner plot showing the two-dimensional probability distributions of dT and dP and the histograms posterior probability distribution of both (units are in minutes).

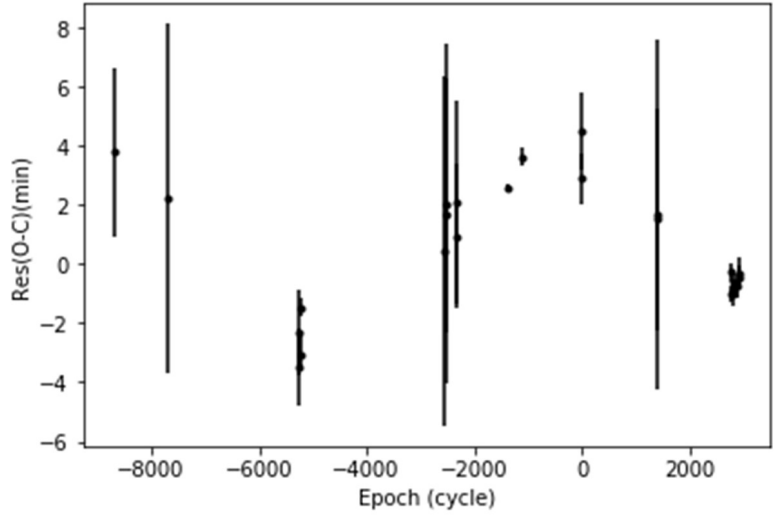

**Fig 4.** The residue O-C diagram of BQ Ari.

The cyclic trend in the residuals of the O-C diagram in this system can be caused either by the Light-Time Effect (LiTE) due to the existence of a third body or the magnetic activity cycles of the system (Pi et al. 2017). We investigated each of these possible causes to explain the variations in the residuals as models. At first, we assumed the LiTE next to the quadratic trend and obtained the parameters of LiTE induced by the presence of a potential third component in the system (Table 6). They were derived with the GA and the MCMC approach using the OCFit code. The GA removes the necessity of any input values of the model's parameters. Final values of them together with their statistically significant uncertainties were obtained using MCMC fitting. The combination of these two algorithms allowed us to analyze the exact physical model of the observed variations (Gajdoš and Parimucha 2019). The number of generations and the size of one generation were both regarded as 2500 for parameters of GA. Also, we employed 100000 iterations and burn-in=5000 for the O-C diagram in our MCMC runs.

The sinusoidal term suggests a periodic change with a period of 6.28 years and an amplitude of 3.37 minutes. We used Irwin (1952) for the analytical formula for O-C changes as a result of the LiTE. Computations yield a large mass function of $f(m_3) = 0.002\ M_\odot$ for the suspected third companion.

To evaluate the quality of the statistical model and to compare models, in addition to $\chi^2$ and $\chi^2_{red} = \frac{\chi^2}{n-g}$ statistics ($n$ is the number of data points in the fit and $g$ is the number of fit-parameters), we employed the Bayesian Information Criterion (BIC) defined as

$$BIC = \chi^2 + k\ ln\ (n) \qquad (3)$$

where $k$ is the number of variable parameters in the model fit (Neath & Cavanaugh 2012) that the better model has less BIC and $\chi^2_{red}\ near\ to\ 1$.

**Table 6.** Results from the LiTE + Quadratic model for suspected third body in BQ Ari.

| Parameter | Value | Error |
|---|---|---|
| $P_3$ [days] | 2294.98 | 275.96 |
| $P_3$ [years] | 6.28 | 0.75 |
| $K_3$ [minutes] | 3.369 | 1.535 |
| $f(m_3)$ [M$_\odot$] | 0.002 | 0.002 |
| $\chi^2$ | 57.29 | |
| $\chi^2_{red}$ | 3.37 | |
| BIC | 83.04 | |

Alternatively, the period's changes could be explained by quadratic + magnetic activity (Applegate 1992). The large scale orbital period variation of BQ Ari might have also been caused by a magnetic activity cycle and/or spot-induced modulations in the light curves which we assumed to be a sinusoidal cycle. The light curve solutions require a huge cold spot due to the differences in the light curves' maxima, indicating a significant magnetic activity present in this binary system. This trend reveals a cyclic oscillation with an amplitude of $A = 3.37$ minutes shown in Figure 5.

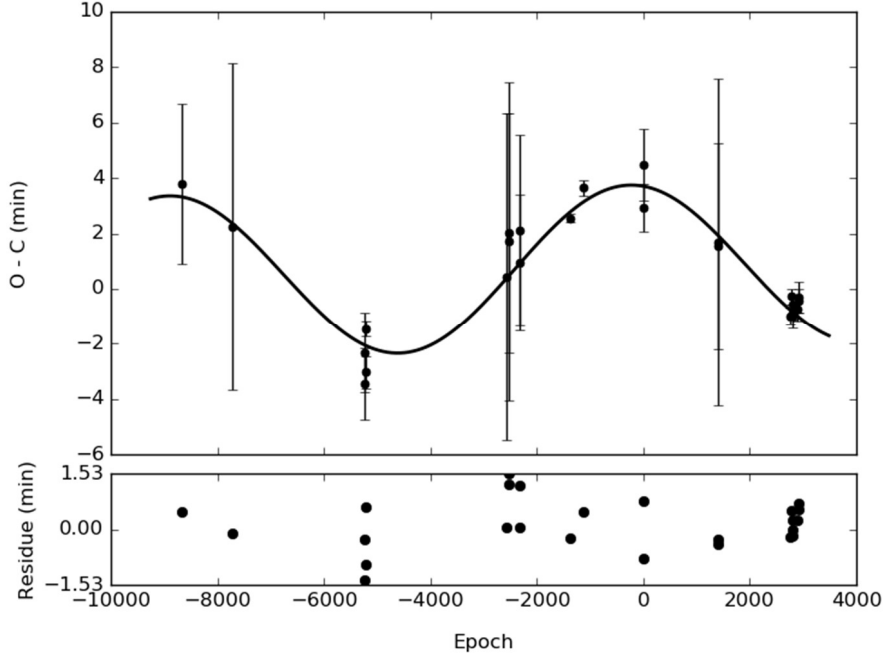

**Fig 5.** The O-C diagram of BQ Ari with the cyclic trend in the data points induced by the magnetic activity cycle.

The results of the fit applied by the OCFit code are provided in Table 7. We measured $B = 18.4\ kG$ as the mean subsurface magnetic field of the cooler component.

**Table 7.** Results from the magnetic activity model of BQ Ari.

| Parameter | Value | Error |
|---|---|---|
| $T[days]$ | 2445.39 | 295.81 |
| $T[years]$ | 6.69 | 0.80 |
| $A[minutes]$ | 2.94 | 0.49 |
| $\chi^2$ | 23.95 | |
| $\chi^2_{red}$ | 1.40 | |
| BIC | 49.70 | |

## 6. CONCLUSION

We presented the first light curve solution and orbital variation analysis of the contact binary system BQ Ari within this study. The photometric observations of BQ Ari were carried out during four nights of observations using $BVR$ filters. We determined its photometric and geometric elements and their uncertainties using the W-D code combined with the MC simulation. This approach gives especially reliable solutions for binary systems with total eclipses. Our results suggest that BQ Ari is a contact binary with a mass ratio of $0.548 \pm 0.019$ and a fillout factor of $24 \pm 0.8\%$, and an inclination of $85.09 \pm 0.45$. Furthermore, there is no radial-velocity information currently available for this binary system, so we were forced to estimate the binary system's absolute parameters based on the Gaia EDR3 parallax.

We suggest a new ephemeris to determine the times of the primary minima for future observations. Since there is a sinusoidal trend in the residuals of the O-C diagram, we provided the alternative solutions; a "LiTE + Quadratic" model and a "Magnetic activity + Quadratic" model. We attempted to explain the residuals of the linear fit in the O-C diagram with these two models, and the best model resulted in $\chi^2_{red}$= 1.4 and with less BIC for the "Magnetic activity + Quadratic" model.

The O'Connell effect can be recognized clearly in the light curves from our observations (Table 5). We also found a light curve in the B.R.N.O. from the year 2017. This was observed by Školník using the Digital Single-Lens Reflex (DSLR) method with a Clear filter. We calculated the difference between the maxima level of this light curve and found a value of $MaxI - MaxII = -0.038$. Although the quality of the light curve is not ideal, the difference between the light levels of the quadrature orbital phases is quite noticeable. On the other hand, light curve solutions required a huge cold spot accounting for the O'Connell effect and the observed light curve asymmetries which hints at a significant magnetic activity in this binary system. Therefore, fitting the orbital period change with a magnetic activity assumption (cyclic) is more probable. It is suggested that there is a longer magnetic cycle, perhaps far longer than our observations. Additional studies should be made that might reveal the nature of orbital period variations in it. Hence these models can be considered as speculations for future reference.

**AVAILABILITY**

The Python codes used in this study will be available at:
https://github.com/Somayeh91/BQ_Ari_analysis/
The data underlying this article will be shared at reasonable request to the corresponding author.

**ACKNOWLEDGMENTS**

This manuscript was prepared by cooperation between the International Occultation Timing Association Middle East section (IOTA/ME) and Çukurova University of Adana, Adana, Turkey. This group activity occurred during the IOTA/ME 2nd Workshop on Photometric Study of Binary Systems and Exoplanet Transits held at Çukurova University, Adana, Turkey, from 4-7 February 2020. We also give special thanks to Prof. Mehmet Emin Özel for his cooperation in this workshop. OB thanks TÜBİTAK for their support with project 118F042 that made his participation possible. We thank TÜBİTAK National Observatory for its support in providing the CCD to UZAYMER. Also, great thanks to Paul D. Maley for making some corrections in the text.